\documentclass[%
 reprint,
 superscriptaddress,
 amsmath,amssymb,
 aps,
 pra,
]{revtex4-2}

\usepackage{graphicx}
\usepackage{dcolumn}
\usepackage{bm}
\usepackage{color}
\usepackage{hyperref}
\hypersetup{colorlinks, linkcolor=blue}

\begin{document}

\title{Optical Solitary Wavelets} 

\author{Oliver Melchert}
\email{melchert@iqo.uni-hannover.de}

\author{Ayhan Demircan}%

\affiliation{%
Institute of Quantum Optics, Leibniz Universität Hannover, Welfengarten 1, 30167 Hannover, Germany
}%
\affiliation{%
Cluster of Excellence PhoenixD, Leibniz Universität Hannover, Welfengarten 1A, 30167 Hannover, Germany
}%

\date{\today}

\begin{abstract}
We investigate wavelet-like localized solutions in nonlinear waveguides, enabled by complementary propagation constants embedded in domains of anomalous dispersion.
They are carrier-envelope-phase stable and independent of fine details of the dispersion profile. 
Their spectra extend over vast ranges and they are also robust against nonlinear perturbations, providing properties highly demanded in ultrafast science.
In the limits of low and high energies, approximate representations are given by modified Morlet wavelets and Gaussian pulses, respectively.
We demonstrate their special features by exploiting general principles of quantum scattering theory and solid-state physics, realizing interlocked chains of wavelets analog to photonic time crystals.
\end{abstract}

\maketitle

\tableofcontents

\section{Introduction}
Ultrafast science explores dynamics and phenomena on ever decreasing time-scales, highly interesting from a fundamental point of view and applications.
Presently moving towards the realm of attosecond physics \cite{Krausz:RMP:2009}, it charts the way for a future development of novel applications in optics \cite{Corkum:NP:2007,Krausz:NP:2014}, such as attosecond light sources and ultrafast lasers.
It therefore has to cope with innate difficulties related to the generation of ultrashort pulses, their manipulation and even just undisturbed transport. 
Optical solitons \cite{Hasegawa:APL:1973,*Hasegawa:APL:1973_2,Mollenauer:PRL:1980}, which allow for and and keep offering new and exciting perspectives~\cite{Redondo:NP:2023}, possibly play an important role in realizing these upcoming developments.
So far, considering short pulses, soliton based control and manipulation is rather vulnerable: they require a delicate balance of linear and nonlinear effects over wide ranges \cite{Dudley:RMP:2006,Agrawal:BOOK:2019,Babushkin:LSA:2017}. 
New types of solitary waves (SWs), distinct from those of standard systems, have been shown recently in several works \cite{Karlsson:OC:1994,Kruglov:PRA:2018,Redondo:NC:2016,Tam:OL:2019,Widjaja:PRA:2021,Tsoy:PRA:2024,Kruglov:PRA:2020,Triki:PRE:2020,Tam:PRA:2020,Lourdesamy:NP:2021,Runge:PRR:2021}.
Here we demonstrate a concept for SWs that exhibit the prerequisites for ultrafast applications in a natural way. 
Carrier-envelop-phase (CEP) stability is an inherent characteristic of these objects.
They can feature highly coherent supercontinuum spectra, unperturbed by special profiles of the linear waveguide dispersion, and can defy the Raman induced self-frequency shift (SFS).
They also provide further unique properties which can be exploited to build more complex states of light, such as, e.g., a new kind of photonic time-crystal.
These solutions share structural similarities with Morlet-wavelets, commonly employed in signal processing applications ranging from solid-state physics \cite{Hase:Nature:2003} to cosmology \cite{LIGO:PRL:2016,Tohfa:PRL:2024}.
Key prerequisite for the existence of these SWs is a complementary propagation constant, embedded in regions with anomalous group-velocity-dispersion (GVD).
For such systems, new effects such as spectral tunneling \cite{Tsoy:PRA:2007,Serkin:EL:1993}, two-frequency soliton molecules and photonic meta-atoms \cite{Melchert:PRL:2019,Willms:PRA:2022,Melchert:OL:2021,Melchert:OPTIK:2023} have been demonstrated.

\section{Model}
\begin{figure}[t!]
\includegraphics[width=\linewidth]{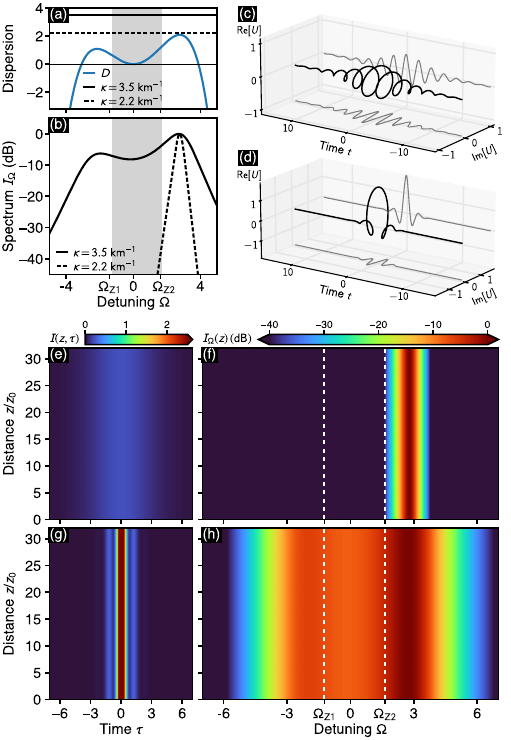}
\caption{Wavelet-like solutions of Eq.~(\ref{eq:HONSE}). 
(a) $M$-shaped dispersion relation $D$. Horizontal lines 
indicate wavenumbers of two selected solutions. Shaded area exhibits normal dispersion.
(b-d) Localized solutions obtained using the SRM.
(b) Spectra $I_\Omega=|U_\Omega|^2/\max(|U_\Omega|^2)$ of the solutions indicated in (a).
(c) Complex-valued field for $\kappa=2.2~\mathrm{km^{-1}}$, and, (d) for $\kappa=3.5~\mathrm{km^{-1}}$.
(e-f) Pulse propagation simulation for the solution with $\kappa=2.2~\mathrm{km^{-1}}$. (e) Intensity $I(z,\tau)=|A(z,\tau)|^2$. (f) Scaled spectrum $I_\Omega(z)=|A_\Omega(z)|^2/\max(|A_\Omega(0)|^2)$. Vertical dashed lines indicate zero-dispersion points.
Distance scaled by $z_0=2\pi/\kappa$. 
(g-h) Same as (e-f) for $\kappa=3.5~\mathrm{km^{-1}}$.
}
\label{fig:01}
\end{figure}

We focus on propagation constants with alternate domains of anomalous and normal dispersion that can be realized in slot-waveguides \cite{Zhang:OE:2012}, photonic crystal fibers \cite{Willms:PTL:2023}, and hollow-core fibers \cite{Zeisberger:SR:2017,*Debord:Fibers:2019}.
A simple model of this type has positive second-order dispersion (2OD), third-order dispersion (3OD), and especially negative fourth-order dispersion (4OD), governed by the higher-order nonlinear Schrödinger equation (HONSE) 
\begin{align}
i \partial_z A =  \frac{\beta_2}{2} \partial_\tau^2 A + i\frac{\beta_3}{6}\partial_\tau^3 A - \frac{\beta_4}{24} \partial_\tau^4 A - \gamma |A|^2 A, \label{eq:HONSE}
\end{align}
for an envelope $A\equiv A(z,\tau)$, propagation distance $z$, and retarded time $\tau=t-\beta_1 z$. $\gamma$ is a nonlinear parameter, and $\beta_1$, $\beta_2$, $\beta_3$, and $\beta_4$ are the parameters specifying the group delay, 2OD, 3OD, and 4OD, respectively. 
For $\beta_2>0$ and $\beta_4<0$, Eq.~(\ref{eq:HONSE}) features a $M$-shaped dispersion relation $D(\Omega)=\tfrac{\beta_2}{2}\Omega^2 + \tfrac{\beta_3}{6}\Omega^3 + \tfrac{\beta_4}{24}\Omega^4$ [see Fig.~\ref{fig:01}(a)], with $\beta_3\neq 0$ breaking the symmetry about the detuning $\Omega=0$.
It exhibits two zero-dispersion points (ZDPs) at 
$\Omega_{{\rm{Z1}},{\rm{Z2}}}=(-\beta_3 \pm \sqrt{\beta_3^2 - 2\beta_2\beta_4})/\beta_4$, delimiting a domain of normal dispersion that separates two domains of anomalous dispersion within which quasi-group-velocity matched co-propagation of optical pulses can be achieved \cite{Melchert:PRL:2019,Melchert:OPTIK:2023}.
Its local maxima are located at $\Omega_{\rm{M1},\rm{M2}} = (-\tfrac{3}{2}\beta_3 \pm \sqrt{  \tfrac{9}{4}\beta_3^2 - 6\beta_2 \beta_4 })/\beta_4$.
%
%
Previously, in systems similar to Eq.~(\ref{eq:HONSE}), SWs have been demonstrated \cite{Melchert:PRL:2019,Melchert:OL:2021,Melchert:OL:2023}, and their interaction dynamics have been studied \cite{Willms:PRA:2022,Oreshnikov:PRA:2022,Melchert:NJP:2023}. 
Specifically, for $\beta_3=0$, generalized dispersion Kerr solitons (GDKSs) where identified, and a closed-form solution of their meta-envelope (ME) was obtained in the long-pulse limit \cite{Tam:PRA:2020}. 
Such objects have been observed in mode-locked laser cavities \cite{Lourdesamy:NP:2021,Mao:NC:2021,Cui:PRL:2023,Widjaja:OL:24}, and extensions to multiple domains of anomalous dispersion have been addressed \cite{Willms:PRA:2022,Lourdesamy:JOSAB:2023}.
In the long-pulse limit, a simplified model based on two incoherently coupled nonlinear Schrödinger equations (NSEs) applies, admitting two-frequency soliton pairs (TFSPs) as solutions \cite{Melchert:OL:2021}.
Their total field reads $A(z,\tau)=U(\tau)\,e^{i\kappa z}$ with real-valued amplitude \cite{Melchert:OL:2021}
\begin{align}
 U(\tau)=\sqrt{\frac{8 \beta_2}{3\gamma t_0^2}}\,{\mathrm{sech}\left(\frac{\tau}{t_0}\right)}\,\cos\left(\sqrt{\frac{6\beta_2}{|\beta_4|}} \tau\right), \label{eq:U}
\end{align}
where $\kappa= \kappa_0 + \beta_2/t_0^2$  with $\kappa_0 \equiv \max(D) = 3 \beta_2^2/(2|\beta_4|)$ is the nonlinear wavenumber,  $t_0$ is the duration of the secant-shaped envelope. 
This amplitude contains the ME first determined by Tam {\emph{et al.}} \cite{Tam:PRA:2020,Tam:PRA:2020:cmp}.
Let us note that Eq.~(\ref{eq:U}) has the form of the modified Morlet wavelet, introduced to analyze optical pulses in Ref.~\cite{Landolsi:OFT:2001}. 
Below we thus adopt the term \emph{wavelet} to refer to solutions of Eq.~(\ref{eq:HONSE}) for $\beta_2>0$, general $\beta_3$, and $\beta_4<0$.

\section{Results}
Here we generalize previous studies \cite{Melchert:PRL:2019,Tam:PRA:2020,Willms:OL:2023,Tsoy:PRA:2024} and numerically calculate SWs of Eq.~(\ref{eq:HONSE}) using a custom spectral renormalization method (SRM; see Supplemental Material \cite{suppMat}).
It admits an Ansatz $A(z,\tau)=U(\tau)\,e^{i\kappa z}$ with complex valued $U$, and provides previously unknown solutions, especially for nonzero $\beta_3$. 
For convenience we set parameters to $\gamma=1~\mathrm{W^{-1}/km}$, $\beta_2=1~\mathrm{ps^2/km}$, $\beta_3=0.2~\mathrm{ps^3/km}$, and $\beta_4=-1~\mathrm{ps^4/km}$, for which 
$\Omega_{\rm{M1},\rm{M2}}\approx(-2.17,2.77)~\mathrm{rad/ps}$, and $\kappa_0\approx 2.1~\mathrm{km^{-1}}$.
We consider an example with parameters tailored to ultrashort pulse propagation in presence of self-steepening and Raman effect further below.

\begin{figure*}[t!]
\includegraphics[width=\linewidth]{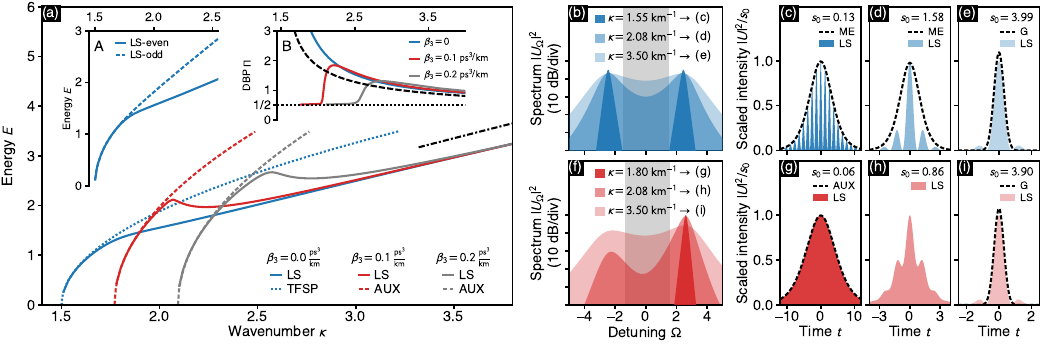}
\caption{Properties of wavelet-like localized solutions.
(a) Energy-wavenumber diagram. 
Solid lines (labeled LS) indicate energies for selected values of $\beta_3$.
Short dashed line (TFSP) indicates energy of TFSPs for $\beta_3=0$.
Dashed lines (AUX) indicate energy of auxiliary solitons. 
Dashed dotted line shows energy of approximate Gaussian solutions.
Inset A compares wavelet energies for even-parity (LS-even) and odd-parity (LS-odd) for $\beta_3=0$.
Inset B shows the duration-bandwidth product (DBP).
(b-e) Results for $\beta_3=0$ and selected wavenumbers. (b) Spectra. Grey shaded area indicates region of normal dispersion. (c) Intensity for $\kappa=1.55~\mathrm{km^{-1}}$. Dashed line indicates the ME of Ref.~\cite{Tam:PRA:2020}. (d) Results for $\kappa=2.08~\mathrm{km^{-1}}$. (e) Results for $\kappa=3.5~\mathrm{km^{-1}}$. Dashed line indicates approximate Gaussian solution.
(f-i) Same as (b-e) for $\beta_3=0.1~\mathrm{ps^3/km}$ and wavenumbers $\kappa=1.8$, $2.08$, and $3.5~\mathrm{km^{-1}}$. Dashed line in (g) indicates intensity of an auxiliary soliton for $\beta_2^\prime = -2.12\,\mathrm{ps^2/km}$.
}
\label{fig:02}
\end{figure*}

\subsection{Wavelet-like solutions for Eq.~(\ref{eq:HONSE})}
Considering the small wavenumber $\kappa=2.2~\mathrm{km^{-1}}$ ($\kappa\approx \kappa_0$) [Fig.~\ref{fig:01}(a)], the SRM solution has a narrow spectrum localized at $\Omega\approx 2.77~\mathrm{rad/ps}$, i.e.\ right at the larger lobe of the dispersion profile at  $\Omega_{\rm{M2}}$ [Fig.~\ref{fig:01}(b)].
Its time-domain representation is complex valued [Fig.~\ref{fig:01}(c)], with even real part, odd imaginary part, and many cycles within the wavelet. 
For the larger wavenumber $\kappa=3.5~\mathrm{km^{-1}}$ [Fig.~\ref{fig:01}(a)], the SRM solution exhibits a more intricate structure: its spectrum stretches over the entire domain of normal dispersion, with a dominant peak at $\Omega\approx 2.71~\mathrm{rad/ps}$ and a lesser peak at $\Omega\approx -1.97~\mathrm{rad/ps}$ [Fig.~\ref{fig:01}(b)], both shifted away from the local maxima of $D$; its time-domain wavelet is characterized by few cycles only [Fig.~\ref{fig:01}(d)].
Both wavelets have real-valued spectral amplitudes.
Their stationarity is demonstrated in Figs.~\ref{fig:01}(e-h) by pulse propagation simulations in terms of the conservation quantity error method \cite{Heidt:JLT:2009,Melchert:CPC:2022}, using the energy $E(z)=\int |A(z,\tau)|^2~{\rm{d}}\tau$ to guide stepsize selection.
The fringes within the wavelets intensity profile remain stationary, implying that their group velocity matches their phase velocity.
Evaluating this condition for the dispersion profile of linear waves as $\partial_\Omega D(\Omega) = D(\Omega)/\Omega$ yields the three roots $\Omega_0=0$, and $\Omega_{\pm} = (-2\beta_3 \pm \sqrt{ 4\beta_3^2 - 6 \beta_2\beta_4 })/\beta_4$.
The frequencies $\Omega_+ \approx 2.88~\mathrm{rad/ps}$ and $\Omega_- \approx -2.08~\mathrm{rad/ps}$ predict the peak loci of the latter wavelet quite well.
This shows that CEP stability is generically given for these solutions.
In support of this conclusion, when generating large wavenumber wavelets through collisions in pulse propagation simulations \cite{Willms:OL:2023}, the unique property of CEP sets in automatically. 
The fringe spacing in the time-domain is given by $\Delta t = 2\pi/|\Omega_{+}-\Omega_{-}|$, with $\Delta t \approx \sqrt{|\beta_4|/\beta_2}$ for small $|\beta_3|$. 
Thus, considering $\beta_2=1~\mathrm{ps^2/km}$, a value of $\beta_4=-10^{-6}~\mathrm{ps^4/km}$ yields a central wavelet peak of extent $\Delta t \approx 1~\mathrm{fs}$.

\subsection{Approximate solutions for small $\kappa$}
For vanishing $\beta_3$ ($\kappa_0=1.5~\mathrm{km^{-1}}$) 
and in the long-pulse limit $\kappa\approx \kappa_0$,  where approximate solutions are given by TFSPs \cite{Melchert:OL:2021}, we expect the wavelets
to approach Eq.~(\ref{eq:U}).
As evident from Fig.~\ref{fig:02}(a), the energy difference indeed vanishes as $\kappa\to \kappa_0$. 
Let us note that the incoherent coupling approach of Ref.~\cite{Melchert:OL:2021} also admits TFSPs with odd-parity, given by Eq.~(\ref{eq:U}) with $\sin$ in place of $\cos$.
Their energies agree for $\kappa\to \kappa_0$, where $E_{\rm{TFSP}}=\tfrac{8}{3\gamma}\sqrt{\beta_2 (\kappa-\kappa_0)}$  (see Supplemental Material \cite{suppMat}).
For any given value of $\kappa>\kappa_0$, however, the even-parity wavelet obtained via the SRM exhibits smaller energy than the corresponding odd-parity wavelet [inset A of Fig.~\ref{fig:02}(a)]. 
This suggest the former to be dynamically stable, while the latter is not.
We find this confirmed by pulse propagation simulations for odd-parity wavelets: small perturbations amplify upon propagation, causing it to decay into an even-parity wavelet and residual radiation (see Supplemental Material \cite{suppMat}). 
Examples of wavelets for $\beta_3=0$ are shown in Figs.~\ref{fig:02}(b-e). 
For $\kappa\approx \kappa_0$ [Fig.~\ref{fig:02}(c)], the ME of Ref.~\cite{Tam:OL:2019} fits the wavelet obtained in terms of the SRM very well.
For increasing $\kappa$, an intense central peak protrudes over secondary sidelobes, and the  ME does not provide a good fit any more [Fig.~\ref{fig:02}(d)].
For nonzero $\beta_3$, in contrast, the wavelets spectra lack the symmetry about $\Omega=0$  [Figs.~\ref{fig:02}(f)].
While TFSPs are not suited to describe the long-pulse (low energy) limit in this case, the narrow spectra of wavelets for $\kappa \approx \kappa_0$ [Fig.~\ref{fig:01}(b)] suggest a simple auxiliary model:
a standard NSE with the value of GVD taken at maximum dispersion, i.e.\ $\beta_2^\prime = \partial_\Omega^2 D(\Omega)|_{\Omega=\Omega_{\rm{M2}}}$, describes this limit appropriately.
The corresponding fundamental solitons are in excellent agreement with our SRM solutions [Fig.~\ref{fig:02}(g)], and their energy $E_{\rm{AUX}}=\tfrac{2}{\gamma}\sqrt{2|\beta_2^\prime|(\kappa-\kappa_0)}$ matches our numerical results very well [Fig.~\ref{fig:02}(a)].
For increasing $\kappa$, while the primary peak in the spectrum vastly exceeds the secondary peak, a wavelet resembles a photonic meta-atom, i.e.\ a strong pulse trapping a weak pulse using an incoherent XPM-induced binding mechanism \cite{Melchert:PRL:2019,Melchert:OPTIK:2023}. Its energy is still governed by $E_{\rm{AUX}}$.
As the trapped pulse grows stronger, their mutual interaction favors the formation of an inseparable unit with reduced energy. 
As evident from Fig.~\ref{fig:02}(a), especially for $\beta_3\neq 0$, 
the usual unambiguous energy-wavenumber relation of solitons is lost, 
now allowing for energy-degenerate solutions at distinct wavenumbers.
This change in structure affects the localization of energy within a wavelet, expressed by their duration-bandwidth product $\Pi$ (DBP; see Supplemental Material \cite{suppMat}).
As evident from Fig.~\ref{fig:02}(a) (inset B), once the nontrivial structure of wavelets is manifested, $\Pi$ greatly exceeds the theoretical minimal value $\Pi=1/2$ of Gaussian pulses (dotted line). 
Specifically, for $\beta_3=0$ and on basis of Eq.~(\ref{eq:U}) we estimate $\Pi\approx \pi \sqrt{\beta_2^2/[6\,|\beta_4|\,(\kappa-\kappa_0)]}$ (dashed line), extending previous results \cite{Tam:PRA:2020}.
For $\beta_3\neq 0$ and $\kappa \to \kappa_0$, it approaches the value $\Pi\approx 0.525$ for solitons of the standard NSE.

\subsection{Approximate solutions for large $\kappa$}
Let us note that for large wavenumbers, the energy [Fig.~\ref{fig:02}(a)], DBP, and intensity profile [Figs.~\ref{fig:02}(e,i)] of a wavelet are virtually independent of $\beta_3$.
We might intuitively argue that, in this limit, a solution only ``sees'' the large-scale structure of the dispersion profile, ultimately determined by $\beta_4<0$, but is insensitive to small-scale perturbations caused by nonzero $\beta_3$. 
In this limit, a simplified description in terms of their pronounced central peak seems legitimate.
Constructing approximate solutions using a Gaussian Ansatz $A_{\rm{G}}(z,\tau)=A_0 \exp\{-(\tau/t_0)^2 + i\kappa z\}$, we find 
\begin{subequations}\label{eqs:G}
\begin{align}
t_0^2 &= -\frac{|\beta_2|}{2 \gamma A_0^2} + \sqrt{ \frac{|\beta_2|^2}{4 \gamma^2 A_0^4} + \frac{|\beta_4|}{\gamma A_0^2}},~\text{and},\label{eq:G01}\\
\kappa &= \frac{|\beta_2|}{2 t_0^2} + \frac{\gamma A_0^2}{2}, \label{eq:G02}
\end{align}
\end{subequations}
 relating pulse amplitude and duration [Eq.~(\ref{eq:G01})], and linking the envelope parameters to the pulse wavenumber [Eq.~(\ref{eq:G02})] (see Supplemental Material \cite{suppMat}).
As evident from Figs.~\ref{fig:02}(e,i), already at $\kappa=3.5~\mathrm{km^{-1}}$ and independent of $\beta_3$, the resulting parameters $A_0\approx 2.09~\mathrm{\sqrt{W}}$ and $t_0\approx 0.61~\mathrm{ps}$ approximate the exact solutions reasonably well.  
Also, their energy $E_{\rm{G}}=\sqrt{\pi/2} \,t_0 A_0^2$ matches the actual energy quite well [Fig.~\ref{fig:02}(a)].
We emphasize that for $\kappa\gg\kappa_0$ (high energy), i.e.\ for large peak amplitudes $A_0$,
 Eq.~(\ref{eqs:G}) approaches the PQS limit $t_0 = [|\beta_4|/(\gamma A_0^2)]^{1/4}$ \cite{Redondo:NC:2016}.
This finding is consistent with earlier results for GDKSs \cite{Tam:PRA:2020}, and recent results on generic PQSs \cite{Tsoy:PRA:2024}.

\begin{figure}[b!]
\includegraphics[width=\linewidth]{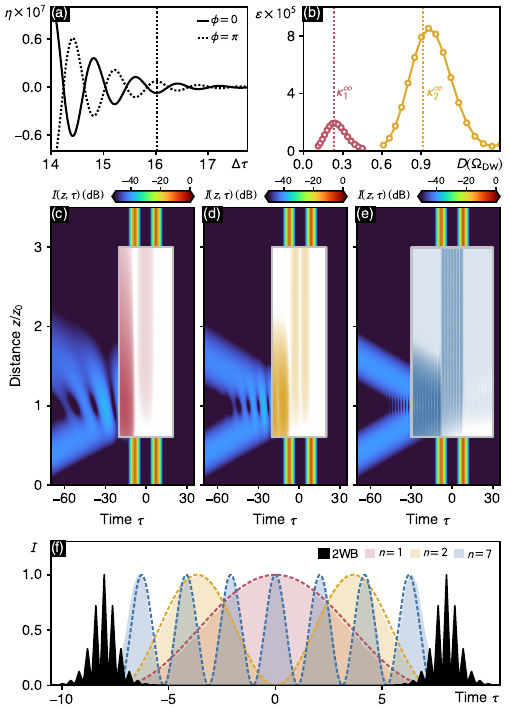}
\caption{Scattering of dispersive waves (DWs) at an interlocked two-wavelet barrier (2WB).
(a) Wavelet-wavelet interaction. Vertical dashed line indicates local minimum at  $\Delta\tau \approx 16.02~\mathrm{ps}$ ($\phi=0$).
(b) Fractional DW energy $\epsilon$ transferred to a 2WB resonance. Vertical dashed lines indicate wavenumbers $\kappa_1^{\infty} \approx 0.19~\mathrm{km^{-1}}$ and $\kappa_2^{\infty}\approx 0.92~\mathrm{km^{-1}}$ of infinite square well. 
(c-e) Scattering of DWs at different detunings $\Omega_{\rm{DW}}$ off the 2WB. Propagation distance scaled by $z_0=\tau_0/\beta_1(\Omega_{\rm{DW}})$ with $\tau_0=80~\mathrm{ps}$. Insets show filtered view of the incident DW and the resulting resonance.
(c) $\Omega_{\rm{DW}}=0.22~\mathrm{rad/ps}$,
(d) $\Omega_{\rm{DW}}=0.44~\mathrm{rad/ps}$, and,
(e) $\Omega_{\rm{DW}}=1.49~\mathrm{rad/ps}$.
(f) Scaled intensities $I(\tau)=|A(z,\tau)|^2/\max(|A(z,\tau)|^2)$ of the 2WB, and selected resonant states of order $n$, resulting from numerical simulations at $z/z_0=3$. Dashed lines indicate bound states of infinite square well.
}
\label{fig:04}
\end{figure}

\subsection{One-dimensional potentials and crystals} 
Solitons are exact solutions only when they are well separated from other solitons.
In the standard NSE, their mutual interaction causes a net attraction or repulsion \cite{Gordon:OL:1983,Mitschke:OL:1987}.
Dispersion engineering can be used to stabilize their distance utilizing a phase relation, enabling soliton molecules \cite{Stratmann:PRL:2005,Hause:PRA:2008}. 
Here, the challenge of stabilizing two or more wavelets relative to each other can be met by exploiting a ``long-range'' binding mechanism for SWs with oscillatory tails \cite{Malomed:PRA:1991,Malomed:PRE:1993}.
Viewed as function of time lag $\Delta \tau$, and determined by their oscillatory temporal profile and complex phase structure, the wavelet-wavelet interaction $H_{\rm{int}}(\Delta \tau,\phi)=-\gamma\,\int |U(\tau) + U(\tau\!+\!\Delta \tau)\,e^{i\phi}|^4\,{\mathrm{d}}\tau$ exhibits periodic local minima.
An example of the interaction parameter $\eta(\Delta\tau,\phi)\equiv 1-H_{\rm{int}}(\Delta,\phi)/H_{\rm{int}}(\Delta\tau\!\to\!\infty, \phi)$ for parameters $\beta_2=10~\mathrm{ps^2/km}$, $\beta_4=-1~\mathrm{ps^4/km}$, $\gamma=1~\mathrm{W^{-1}/km}$, and a wavelet $U$ for $\kappa\approx 18~\mathrm{km^{-1}}$ is shown in Fig.~\ref{fig:04}(a).
This can be utilized to construct one-dimensional photonic time crystals with Kerr-comb-like spectra by interlocking adjacent wavelets using their tails (see Supplementary Material \cite{suppMat}).
A minimal example, consisting of two wavelets interlocked at $\Delta \tau\approx 16.02~\mathrm{ps}$ for $\phi=0$, is demonstrated in Fig.~\ref{fig:04}. 
In extension of a soliton, acting as refractive index barrier for a quasi group-velocity matched dispersive wave (DW) in a domain of normal dispersion \cite{Demircan:PRL:2011}, an interlocked bound state of two wavelets forms a double barrier. Analogous to double barriers in 1D quantum scattering \cite{Goff:AJP:2023}, the two-wavelet barrier (2WB) exhibits resonances as function of the incident DW wavenumber $\kappa_n\equiv D(\Omega_{{\rm{DW}},n})$ at certain detunings $\Omega_{{\rm{DW}},n}$.
A simplified model of the 2WB in terms of an infinite square well with extend $\Delta \tau_\infty = 14.7~\mathrm{ps}$, i.e.\ with walls placed at the full-width at half-maximum points of the 2WB, yields bound-state wavenumbers $\kappa_n^\infty= (n\pi)^2 \beta_2/(2 \Delta \tau_{\infty}^2)$ and 
amplitudes $\psi_n^\infty(\tau) = \sqrt{2/\Delta\tau_\infty}\,\sin( n \pi \Delta \tau_{\infty}^{-1} [\tau\!+\!\Delta\tau_\infty/2])$, for $|\tau| \leq \Delta\tau_\infty/2$.
This should allow to describe low-lying resonances reasonably well, predicting a GS wavenumber $\kappa_1^\infty\approx 0.23~\mathrm{km^{-1}}$.
From pulse propagation simulations of the HONSE~(\ref{eq:HONSE}) with initial condition 
$A_0 = A_{\rm{2WB}} + A_{\rm{DW}}$, where $A_{\rm{2WB}}(\tau)=U(\tau+\tfrac{\Delta \tau}{2}) + U(\tau-\tfrac{\Delta\tau}{2})$, and $A_{\rm{DW}}(\tau)=10^{-2}\,U_0\,\exp\{-(\tau-\tau_0)^2/(2 \sigma^2) - i\Omega_{\rm{DW}} \tau\}$ with $U_0=\max(|U(\tau)|)\approx 6.18~\mathrm{\sqrt{W}}$, $\tau_0=-80~\mathrm{ps}$ and $\sigma=20~\mathrm{ps}$, we find that the maximum amount of energy transfer from the incident DW to the lowest-order resonance of the 2WB occurs at $\kappa_1=0.25\,\mathrm{km^{-1}}$ for $\Omega_{{\rm{DW}},1}=0.22~\mathrm{rad/fs}$ [Fig.~\ref{fig:04}(b)].
In this example, the auxiliary square well exhibits 18 bound states with wavenumbers below the 2WB height $V_0=2\gamma U_0^2$.
The resonant excitation of states of order $n=1$, $2$, and $7$, are demonstrated in Figs.~\ref{fig:04}(c-e).
Even the 7th resonance is reproduced satisfactorily in the infinite square well approximation [Fig.~\ref{fig:04}(f)].
While all of these states are metastable, the lowest order resonances exhibit especially large $z$-lifetimes. 
In case of $n=7$, the energy stored in the 2WB decays $\propto \exp(-z/z_{\rm{m}})$ with mean lifetime $z_{\rm{m}}\approx 33\,z_0$ (see Supplemental Material \cite{suppMat}).
Wavelets at larger wavenumbers yield deeper square wells and allow for more resonances.
The loci of the resonances and thus their $z$-lifetimes can be controlled via the time-lag $\Delta \tau$ of the interlocked wavelets. 
We further observe that when sufficient energy gets transferred to a resonance, e.g.\ via multiple scatterings, oscillations of the 2WB along the time coordinate can be excited.
On basis of the results of Refs.~\cite{Melchert:PRL:2019,Oreshnikov:PRA:2022}, for an incident DW located in a domain of anomalous dispersion, the two-wavelet bound state is instead perceived as a two-wavelet potential well.

\subsection{Impact of the Raman effect}
To assess the dynamics in the range of ultrashort pulses, we consider the parameters $\gamma=10^{-10}~\mathrm{W^{-1}/\mu m}$, $\beta_2=0.1~\mathrm{fs^2/\mu m}$, $\beta_3=0.1~\mathrm{fs^3/\mu m}$, and $\beta_4=-0.7~\mathrm{fs^4/\mu m}$, as well as standard models for pulse-self steepening and the Raman effect \cite{Blow:JQE:1989}.
We find that wavelets at large wavenumbers $\kappa$ (FWHM duration $\approx 1~\mathrm{fs}$), irrespective of $\beta_3$, defy the usual Raman induced SFS on short propagation distances (see Supplementary Material \cite{suppMat}). 
On long distances they are, however, prone to spectral tunneling, with phase matching processes facilitating a transfer of energy from higher to lower frequencies \cite{Serkin:EL:1993,Tsoy:PRA:2007}. 
Let us note that this is different from the dynamics of PQS under the impact of the Raman effect alone \cite{Wang:OL:2022}.
Our results do, however, fit in well with earlier findings for two-frequency pulse compounds \cite{Willms:PRA:2022}.

\section{Conclusions} 
The HONSE~(\ref{eq:HONSE}) supports CEP stabilized optical solitary wavelets, supported by a blend of normal and anomalous dispersion regimes.
Their short-pulse limit (large wavenumbers) is independent of the exact dispersion profile and their interaction dynamics enable time-periodic photonic structures which admit resonant excitation of metastable states. 
Our results shed further light on the formation of localized states in nonlinear systems, disclose the impact of the Raman effect on the wavelets propagation dynamics, and extend previous findings on GDKSs \cite{Tam:PRA:2020}, PQSs \cite{Tam:OL:2019}, and generic PQSs \cite{Tsoy:PRA:2024}.
Model systems of this type receive a growing interest \cite{Serkin:EL:1993,Tsoy:PRA:2007,Moille:OL:2018,Melchert:PRL:2019,Tam:OL:2019,Tsoy:PRA:2024,Melchert:OL:2020}, 
and may lead to unexpected advances in generation, transport, and control of ultrashort pulses and their applications, building upon analogies between photonics, quantum mechanics and solid-state physics.

\vskip 0.5cm
\paragraph*{Acknowledgments --} We acknowledge financial support from Germany’s Excellence Strategy within the Cluster of Excellence PhoenixD (Photonics, Optics, and Engineering--Innovation Across Disciplines) (EXC 2122, Project No. 390833453).

\appendix

\section{The spectral renormalization method (SRM)\label{sec:SR}}

Below we detail the numerical calculation of localized wave solutions by a custom spectral renormalization method (SRM). 
In sect.~(\ref{sec:SR:NUM}) we first outline the custom SRM for a propagation equation of the form 
\begin{align}
i \partial_z A =  \frac{\beta_2}{2} \partial_\tau^2 A + i\frac{\beta_3}{6}\partial_\tau^3 A - \frac{\beta_4}{24} \partial_\tau^4 A - \gamma |A|^2 A, \label{eq:HONSE_SM}
\end{align}
i.e.\ Eq.~(1) of the primary document, with pulse envelope $A\equiv A(z,\tau)$, propagation coordinate $z$, and retarded time $\tau=t-\beta_1 z$. In Eq.~(\ref{eq:HONSE_SM}), $\gamma$ is a nonlinear parameter, and $\beta_2$, $\beta_3$, and $\beta_4$ are the parameters specifying second-order dispersion (2OD), third-order dispersion (3OD), and fourth-order dispersion (4OD), respectively.
Section~(\ref{sec:SR:TFSP}) includes an example addressing the limiting case of two-frequency soliton pairs  \cite{Melchert:OL:2021}, for which an analytical expression exists.

\subsection{Numerical procedure\label{sec:SR:NUM}}

The standard SRM \cite{Ablowitz:OL:2005} constitutes a modification of the iterative Petviashvili method \cite{Petviashvili:SJPP:1976,Pelinovsky:PRE:2000}, which has been extended to also apply to systems of coupled equations \cite{Ablowitz:OL:2005}, to propagation equations with periodic nonlinear microstructure  \cite{Fibich:PD:2006,Cole:PRA:2014}, as well as to systems with $z$-dependent dispersion map and $z$-dependent nonlinear coefficient \cite{Musslimani:JOSAB:2004}.  
%
%

%
Here we customize the SRM to determine localized stationary solutions (LSS) of Eq.~(\ref{eq:HONSE}) by assuming an Ansatz of the form $A(z,\tau)=U(\tau)\,\exp(i \kappa z)$ with complex valued field $U(\tau)$, satisfying $|U(\tau)| \to 0$ as $|\tau|\to \infty$, and LSS wavenumber  $\kappa$. 
Following this Ansatz, $U\equiv U(\tau)$ is determined by the nonlinear eigenvalue problem
\begin{align}
\kappa U = \hat{D}(i\partial_\tau) U + \gamma |U|^2 U, \label{eq:NLEP}
\end{align}
where $\hat{D}(i\partial_\tau)= \tfrac{1}{2}\beta_2 (i\partial_\tau)^2 + \tfrac{1}{6} \beta_3 (i\partial_\tau)^3 + \tfrac{1}{24}\beta_4 (i\partial_\tau)^4$ governs the effects of linear dispersion.
Subsequently, we consider positve 2OD $\beta_2>0$,
negative 4OD $\beta_4<0$, and arbitrary values of $\beta_3$.
%
%
Considering the discrete set of angular frequency detunings $\Omega\in
\frac{2\pi}{T}\mathbb{Z}$, corresponding to a periodic temporal domain of extent $T$, the field envelope $U(\tau)$ can be related to the spectral envelope $U_\Omega$ by means of the transform-pair
\begin{subequations} \label{eq:FT}
\begin{align}
&U_\Omega= {\mathsf{F}}\left[U(\tau)\right] =\frac{1}{T} \int_{-T/2}^{T/2} U(\tau)\,e^{i\Omega \tau}~{\rm{d}}\tau, \label{eq:FT_FT}\\
&U(\tau)  = {\mathsf{F}}^{-1}\left[ U_\Omega\right] = \sum_{\Omega} U_\Omega\,e^{-i\Omega \tau}. \label{eq:FT_IFT}
\end{align}
\end{subequations}
Using the identity $(i \partial_\tau)^n\,e^{-i\Omega \tau} = \Omega^n \, e^{-i \Omega \tau} $, we write the frequency-domain representation of the dispersion as $D(\Omega)= \tfrac{1}{2}\beta_2 \Omega^2 + \tfrac{1}{6}\beta_3\Omega^3 + \tfrac{1}{24}\beta_4 \Omega^4$. 
%
%
%
\begin{figure}[t!]
\includegraphics[width=\linewidth]{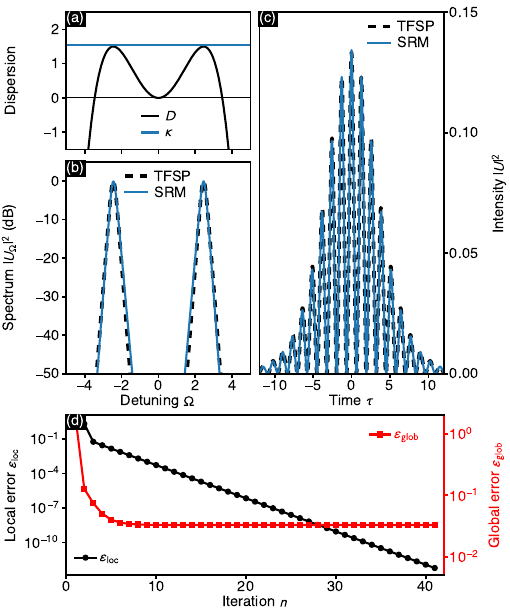}
\caption{Limiting case of two-frequency soliton pairs (TFPCs).
(a) Dispersion profile $D(\Omega)$, and wavenumber $\kappa=1.55~\mathrm{km^{-1}}$ of the sought-for solitary wave solution.
(b-d) Results for a Gaussian trial-function $U_0(\tau)=\exp(-\tau^2/2)$.
(b) Spectrum obtained by the SRM (labeled SRM) and spectrum of the TFSP Eq.~(\ref{eq:TFSP}) (labeled TFSP).
(c) Intensity obtained by the SRM (labeled SRM) and intensity of the approximate solution (labeled TFSP).
(d) Decrease of the local error $\epsilon_{\rm{loc}}$ and the global mismatch $\epsilon_{\rm{glob}}$ upon iteration. 
}
\label{fig:S02}
\end{figure}

Then, starting with a reasonable trial-function $U_0 \equiv U_0(\tau)$ at iteration step $n=0$, the SRM for Eq.~(\ref{eq:NLEP}) consists of the iterative update 
\begin{subequations}\label{eq:SRM_iter}
\begin{align}
    U^\prime &= {\mathsf{F}}^{-1}\left[ \frac{\gamma}{\kappa-D(\Omega)}\,{\mathsf{F}}\left[|U_n|^2 U_n\right]\right],\label{eq:SRM_iter_1}\\
    U_{n+1} &= U^\prime \,\left| \frac{\langle U_n, U_n\rangle}{\langle U_n, U^\prime \rangle}  \right|^{3/2},\label{eq:SRM_iter_2}
\end{align}
\end{subequations}
carried out until the local error $\epsilon_{\mathrm{loc}}\equiv ||U_{n+1}-U_{n}||$ decreases below a given threshold.
In Eq.~(\ref{eq:SRM_iter_2}), \mbox{$\langle a,b \rangle = \int a^*(\tau)\, b(\tau)~{\mathrm{d}}\tau$}, and \mbox{$||a||=\sqrt{\langle a,a\rangle}$}.
In our numerical experiments we interrupt the iteration as soon as \mbox{$\epsilon_{{\mathrm{loc}}} < 10^{-12}$}. 
Let us emphasize that, in order for the SRM to converge to a dynamically stable LSS, the denominator in Eq.~(\ref{eq:SRM_iter_1}) needs to be absent of resonant frequencies $\Omega_{R}$ 
at which $\kappa - D(\Omega_R) = 0$.
For propagation constants with $M$-shaped dispersion curve, this is ensured by limiting the search for LSSs to $\kappa > \kappa_0 \equiv \max\left[D(\Omega)\right]$.

\subsection{Limiting case of two-frequency soliton pair solution\label{sec:SR:TFSP}}

To support our discussion in the primary document, we subsequently demonstrate a use-case for which an approximate solution is available for wavenumbers $\kappa>\kappa_0$ in the limit $\kappa \to \kappa_0$.
%

We subsequently consider $\gamma=1~\mathrm{W^{-1}/km}$, $\beta_2=-1~\mathrm{ps^2/km}$, $\beta_3=0$, and $\beta_4=-1~\mathrm{ps^4/km}$, with $\kappa_0 = 3 \beta_2 /(2 |\beta_4|) = 1.5~\mathrm{km^{-1}}$ attained at the two detuning values $\pm \Omega_{c} = \sqrt{6 \beta_2/|\beta_4|}\approx 2.45~\mathrm{rad/ps}$.
%
We expect that for $\kappa \gtrapprox 1.5~\mathrm{km^{-1}}$, LSSs determined by the SRM are narrowly peaked at the frequency loci $\pm \Omega_{c}$, so that an approximate description of the solution in terms of incoherently coupled NSEs for two separate pulses applies \cite{Melchert:OL:2021}. 
%
An approximate description of the solitary wave solutions of Eq.~(\ref{eq:HONSE}) is then possible in terms of the two-frequency soliton pair (TFSP) solution \cite{Melchert:OL:2021}
\begin{align}
 U(\tau)=\sqrt{\frac{8 \beta_2}{3\gamma t_0^2}}\,{\mathrm{sech}\left(\frac{\tau}{t_0}\right)}\,\cos\left(\sqrt{\frac{6\beta_2}{|\beta_4|}} \tau\right), \label{eq:TFSP}
\end{align}
with wavenumber $\kappa= \kappa_0 + \beta_2/t_0^2$, specifying the fundamental metasoliton first determined by Tam {\emph{et al.}} \cite{Tam:PRA:2020}
(to facilitate comparison with the meta-envelope soliton in Ref.~\cite{Tam:PRA:2020} consider the expansion parameter $\epsilon = \sqrt{2 |\beta_4|/(3\beta_2 t_0^2)}\ll 1$.).
The energy integral $E = \int_{-\infty}^\infty |U(\tau)|^2~{\mathrm{d}\tau}$ for the TFSP solution Eq.~(\ref{eq:TFSP}) can be solved analytically, giving
\begin{align}
E =
\frac{8\beta_2}{3\gamma t_0} \left[1+\pi \Theta_0\,{\mathrm{csch}}(\pi \Theta_0 )\right], \label{eq:E_TFSP}
\end{align}
with $\Theta_0 = \sqrt{6\beta_2/|\beta_4|}\,t_0$, and hyperbolic cosecant ``$\mathrm{csch}$''. Substituting $t_0 = \sqrt{\beta_2/(\kappa-\kappa_0)}$, and taking account of $\mathrm{csch}(\pi \Theta_0)\approx 0$ for $\kappa \to \kappa_0$, yields the estimate 
$E_{\rm{TFSP}}=\tfrac{8}{3\gamma}\sqrt{\beta_2 (\kappa-\kappa_0)}$ used in the primary document.

\begin{figure}[t!]
\includegraphics[width=\linewidth]{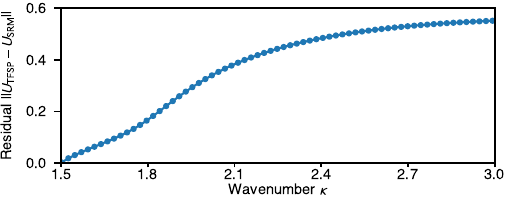}
\caption{Decrease of the residual $||U_{\mathrm{TFSP}}-U_{\mathrm{SRM}}||$ for decreasing wavenumber $\kappa$ (leftmost data point is at $\kappa=1.502~\mathrm{km^{-1}}$).
}
\label{fig:S03}
\end{figure}

In Fig.~\ref{fig:S02} we show the results of the SRM procedure for a solitary wave with wavenumber $\kappa=1.55~\mathrm{km^{-1}}\approx\kappa_0$. 
Let us emphasize that the spectrum and intensity of the SRM solution agree very well with the above TFSP, see Figs.~\ref{fig:S02}(b,c). 
As evident from Fig.~\ref{fig:S02}(d), the local error exhibits an exponential decrease that leads the SRM to converge after 41 iterations.
The global mismatch $\epsilon_{\rm{glob}}=||U_{n+1} - U_{\rm{TFSP}}||$ with respect Eq.~(\ref{eq:TFSP}) settles at a nonzero value of $\epsilon_{\rm{glob}}\approx 0.033$, reinforcing that the TFSP is only an approximate solution.
We further verified that the localized solutions $U_{\mathrm{SRM}}$, obtained in terms of the SRM, approach the TFSP in the limit $\kappa\to \kappa_0$. This is shown in Fig.~\ref{fig:S03}, indicating that the residual $||U_{\mathrm{TFSP}}-U_{\mathrm{SRM}}||$, with $U_{\rm{TFSP}}$ as per Eq.~(\ref{eq:TFSP}) tends to zero as $\kappa\to 1.5~\mathrm{km^{-1}}$, following the discussion of the energy difference in the primary document.

\section{Localized solutions with odd parity}

When the dispersion curve is of the form
 $D(\Omega) = \tfrac{1}{2}\beta_2 \Omega^2 + \tfrac{1}{24} \beta_4 \Omega^4$ with 
$\beta_2>0$ and $\beta_4<0$, 
the localized solutions $U(\tau)$ of Eq.~(\ref{eq:NLEP}) are real-valued and exhibit even parity in the coordinate $\tau$.
As discussed in the primary document, in the TFSP limit $\kappa
\to \kappa_0$, the existence of a pair of TFSP solutions with even and odd parity and equal energy seems plausible.
%
This suggests the existence of stationary solutions for Eq.~(\ref{eq:HONSE}) with even parity (EP) and odd parity (OP) at given wavenumber $\kappa$, which should agree with the TFSP solutions for $\kappa\to \kappa_0$.

\begin{figure}[b!]
\includegraphics[width=\linewidth]{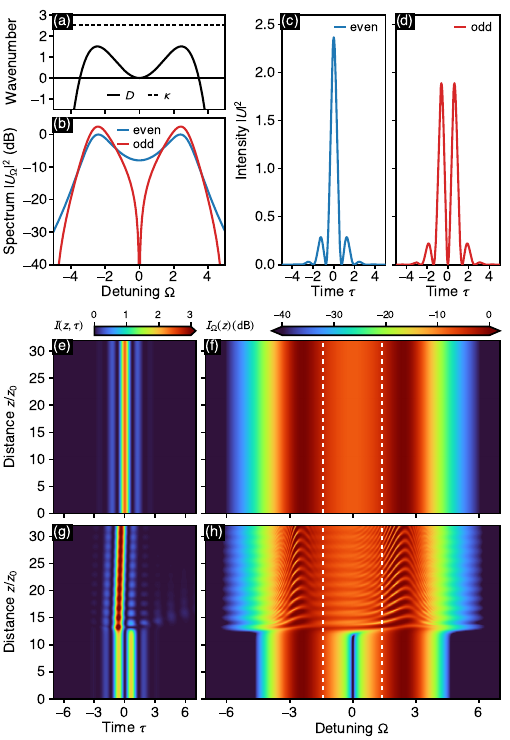}
\caption{SRM solutions with even and odd parity.
(a) Dispersion $D$, and wavenumber $\kappa=2.5~\mathrm{km^{-1}}$ of the sought-for localized wave solutions.
(b) Spectra of even-parity solitary wave (SW) and odd-parity localized wave (LW) obtained by the SRM.
(c) Intensity of the even-parity solution, and, 
(d) intensity of the odd-parity solution.
(e-f) Propagation dynamics of the even-parity solitary wave.
(e) Intensity, and, (f) corresponding spectrum. Dashed lines in (f) indicate zero-dispersion points. Propagation distance is scaled by $z_0=2 \pi /\kappa$.
(g-h) same as (e-f) for the odd-parity localized wave.
}
\label{fig:S04}
\end{figure}

The custom SRM of Sect.~\ref{sec:SR:NUM} can be generalized to determine solutions with odd parity by including the intermediate step 
\begin{align}
  U^\prime \leftarrow U^\prime - {\mathsf{F}}^{-1}\left[ {\mathrm{Re}}\left[{\mathsf{F}}[U^\prime]\right]\right] \label{eq:SRM:iter_1b}
\end{align}
in between Eq.~(\ref{eq:SRM_iter_1}) and Eq.~(\ref{eq:SRM_iter_2}).
This removes the real-valued even-parity part of $U^\prime$ at each iteration step, thus imposing orthogonality between the resulting self-consistent solution and the even-parity LSS $U_{\rm{EP}}$.
Let us note that while the effect of Eq.~(\ref{eq:SRM:iter_1b}) is similar to that of the common Gram-Schmidt orthogonalization procedure
\begin{align}
    U^{\prime} \leftarrow U^\prime - \frac{\langle U^\prime, U_{\rm{EP}} \rangle}{\langle U_{\rm{EP}}, U_{\rm{EP}}\rangle }\,U_{\rm{EP}}, \label{eq:GS_ortho}
\end{align}
it bears the advantage of working with one pulse alone, making it unnecessary to first compute the independent solution $U_{\rm{EP}}$. 
Also, while Eq.~(\ref{eq:GS_ortho}) selects against the specific solution $U_{\mathrm{EP}}$, Eq.~(\ref{eq:SRM:iter_1b}) simply selects for odd parity.

In Figs.~\ref{fig:S04}(a-d) we show exemplary SRM solutions with even and odd parity for $\beta_2=1~\mathrm{ps^2/km}$, $\beta_4=-1~\mathrm{ps^4/km}$, and $\kappa=2.5~\mathrm{km^{-1}}$. 
%
%
As pointed out in the primary document, at a given value of $\kappa > \kappa_0$, the even-parity solution actually exhibits a smaller energy than the odd-parity solution. As a consequence, while the former is dynamically stable [Figs.~\ref{fig:S04}(e,f)], the latter is dynamically unstable [Figs.~\ref{fig:S04}(g,h)]. 
Indeed, as evident from Fig.~\ref{fig:S04}(g), the odd-parity solution ``decays'' into an oscillating even parity solution. 
%
A detailed analysis of the propagation scenario in Fig.~\ref{fig:S04}(g) shows that upon its decay, a fraction of $\approx 0.11$ of the energy $E=2.76~\mathrm{W\,ps}$ of the initial odd-parity solution is lost to dispersive waves, emanating from the localized state. The remaining energy is shifted to left main-lobe and its immediate side-lobes, forming an even parity solution that has a symmetric internal mode excited \cite{Tam:PRA:2020}. The associated amplitude and width oscillations of the
even-parity solutions decay as $z/z_0\to \infty$ ($1/e$-decay length $\approx 6\,z_0$), leaving a SW with energy $E\approx 2.44~\mathrm{W\,ps}$ and wavenumber $\kappa \approx 2.94~\mathrm{km^{-1}}$.

\section{Duration-bandwidth product}

In the main document, the duration-bandwidth product (DBP) is defined as $\Pi=\sigma_t \, \sigma_\Omega$, where $\sigma_t$ (duration) and $\sigma_\Omega$ (bandwidth) are standard deviations describing the wavelets spread in the time domain and frequency domain, respectively.
For a wavelet $U(\tau)$, with $I(\tau)=|U(\tau)|^2$ and $I_\Omega=|U_\Omega|^2$, they are determined by~\cite{Cohen:IEEE:1989}
\begin{subequations}
\begin{align}
\sigma_t^2 = \frac{\int (t-t_c)^2~I(\tau)~{\mathrm{d}}\tau }{\int I(\tau)~{\rm{d}}\tau},~\text{with}~ t_c = \frac{\int t~I(\tau)~{\mathrm{d}}\tau }{\int I(\tau)~{\rm{d}}\tau},
\end{align}
and 
\begin{align}
\sigma_\Omega^2 = \frac{\sum (\Omega-\Omega_c)^2~I_\Omega}{\sum I_\Omega},~\text{with}~ \Omega_c = \frac{\sum \Omega~I_\Omega~}{\sum I_\Omega}.
\end{align}
\end{subequations}
They yield the lower bound $\Pi=1/2$ for Gaussian pulses.

\section{Bound states of wavelets\label{sec:MWB}}

As pointed out in the main document, for a solitary wavelet solution $U$, the wavelet-wavelet interaction $H_{\rm{int}}(\Delta \tau,\phi)=-\gamma\,\int |U(\tau) + U(\tau\!+\!\Delta \tau)\,e^{i\phi}|^4\,{\mathrm{d}}\tau$ exhibits periodic local minima as function of time lag $\Delta \tau$ and mutual phase $\phi$.
This universal ``long-range'' binding mechanism for solitary waves with oscillatory tails was first revealed for spatial solitons in the damped driven NSE \cite{Malomed:PRA:1991,Malomed:PRE:1993}. It was further studied 
for the HONSE with $\beta_2<0$ and $\beta_4<0$ \cite{Buryak:PRE:1995,Akhmediev:OC:1994}, and generalized to temporal cavity solitons in the Lugiato-Lefever equation \cite{Wang:Optica:2017}, and \mbox{pure-quartic solitons \cite{Zeng:AML:2022,Dai:CSF:2022}}.

\subsection{Two-wavelet bound states}
An example for a two-wavelet bound state for parameters $\beta_2=10~\mathrm{ps^2/km}$, $\beta_4=-1~\mathrm{ps^4/km}$, $\gamma=1~\mathrm{W^{-1}/km}$, and a wavelet $U$ for $\kappa\approx 18~\mathrm{km^{-1}}$, interlocked at the local minimum at $\Delta \tau\approx 16.02~\mathrm{ps}$ for $\phi=0$, was demonstrated in Fig.~4 of the main document.
Following the notation of Ref.~\cite{Buryak:PRE:1995}, where bound states of solitons with radiationless oscillating tails for the related, but very different system with $\beta_2<0$ and $\beta_4<0$ where studied, the sets of local minima of $H_{\rm{int}}$ for $\phi=0$ and $\phi=\pi$ indicate the ``symmetric'' and ``antisymmetric'' set of interlocked solutions, respectively. 
Morevover, enumerating the local minima using an integer index $m$, the delay value $\Delta \tau$ corresponds to the $m=7$-th local minimum of 
the symmetric set. 
We may thus refer to the two-wavelet bound state in Fig.~3 of the main text as the symmetric bound state of $7$th order.
Due to the large temporal separation of the wavelets relative to their individual duration, each wavelet overlaps only weakly with the tails of its binding partner.

\begin{figure}[t!]
\includegraphics[width=\linewidth]{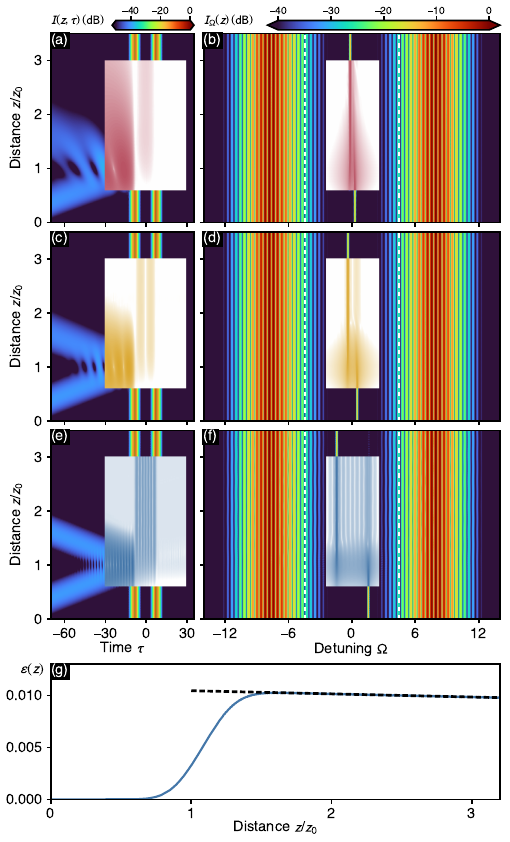}
\caption{Scattering of dispersive waves (DWs) at an interlocked two-wavelet barrier (2WB).
(a-f) Scattering of DWs at different detunings $\Omega_{\rm{DW}}$ off the 2WB. Propagation distance scaled by $z_0=\tau_0/\beta_1(\Omega_{\rm{DW}})$ with $\tau_0=80~\mathrm{ps}$. 
(a-b) Results for $\Omega_{\rm{DW}}=0.22~\mathrm{rad/ps}$. (a) Time-domain propagation dynamics, and, (b) corresponding spectrum. Insets show the associated intensities of the difference field $A^\prime$ (see text for details).
(c-d) Same as (a-b) for $\Omega_{\rm{DW}}=0.44~\mathrm{rad/ps}$, and,
(e-f) Same as (a-b) $\Omega_{\rm{DW}}=1.49~\mathrm{rad/ps}$.
(g) Build-up and decay of the energy fraction $\epsilon$ manifested as the resonance of order $n=7$ [cf.\ subfigures (e-f)], compared to an exponential model $\propto \exp(-z/z_{\rm{m}})$ with mean $z$-lifetime $z_{\rm{m}}\approx 33\, z_0 \approx 230~\mathrm{km}$ ($z_0\approx 6.98~\mathrm{km}$).
}
\label{fig:S05a}
\end{figure}

\paragraph{Uncovering of the resonances in Figs.~3(c-e)}
As discussed in the main text, propagation simulations of Eq.~(\ref{eq:HONSE}) with an initial condition composed of the $7$th order symmetric bound state 
\begin{align}
A_{\rm{2W}}(\tau)=U(\tau+\tfrac{\Delta \tau}{2}) + U(\tau-\tfrac{\Delta\tau}{2}), \label{eq:ic_2W}
\end{align}
and a dispersive wave (DW) in the domain of normal dispersion in the form
\begin{align}
A_{\rm{DW}}(\tau)=10^{-2}\,U_0\,\exp\left\{-\tfrac{(\tau-\tau_0)^2}{2 \sigma^2} - i\Omega_{\rm{DW}} \tau\right\},\label{eq:ic_DW}
\end{align}
 with $U_0=\max(|U(\tau)|)\approx 6.18~\mathrm{\sqrt{W}}$, 
$\tau_0=-80~\mathrm{ps}$ and $\sigma=20~\mathrm{ps}$, 
result in scattering processes in which resonant states of the bound state are excited at certain values of $\Omega_{\rm{DW}}$.
As evident from Eq.~(\ref{eq:ic_DW}) we take the amplitude of the incident DW to be small in comparison to the peak-height $U_0$ of the two-wavelet state. This is done to ensure that the latter remains stationary and is not affected by a frequency shift that accompanies the collision between strong dispersive waves and solitons accross a zero-dispersion point \cite{Demircan:PRL:2011}.
For each considered value of $\Omega_{\rm{DW}}$ we then perform independent pulse propagation simulations for the two initial conditions
\begin{align}
A_0^{\pm}(\tau) = A_{\rm{2W}}(\tau) \pm A_{\rm{DW}}(\tau), \label{eq:ic}
\end{align}
resulting in the fields $A_{\pm}(z,\tau)$.
Technically, this allows us to filter for the ``difference''-field, describing the incident DW and the resonant state, by combining both solutions as $A^\prime(z,\tau)=\tfrac{1}{2}[A_{-}(z,\tau) - A_{+}(z,\tau)]$.
In Fig.~\ref{fig:S05a} we show the propagation dynamics leading to the excitation of the resonances of order $n=1,2$, and $7$, discussed in the main text, along with their spectra. The main panels in Figs.~\ref{fig:S05a}(a,c,e) show the intensities $I[A]=|A(z,\tau)|^2/\max(|A(0,\tau)|^2)$ for the field $A_+$, i.e.\ $I[A_+]$. The insets show the according difference field intensity $I[A^\prime]$. The corresponding spectra are shown in Figs.~\ref{fig:S05a}(b,d,f).
\paragraph{Decay of the resonance of order $n=7$}
While all resonances that can be excited from the outside of the two-wavelet barrier are metastable, they exhibit a rather large $z$-lifetime, see Figs.~
\ref{fig:S05a}(a,c,e). 
Among the shown resonances, only the resonance of order $n=7$ exhibits a noticeable decay to the outside region of the two-wavelet bound state [see the inset of Fig.~\ref{fig:S05a}(e)]. 
To characterize this decay, we determined the fraction of energy 
\begin{align}
\epsilon(z) = \int_{-\Delta \tau/2}^{+\Delta \tau/2} |A^\prime(z,\tau)|^2~{\rm{d}}\tau / \int |A^\prime(0,\tau)|^2~{\rm{d}}\tau, \label{eq:epsilon}
\end{align}
manifested as resonant state. For the $7$th order resonance we found that, following the build up of the resonance in the scaled range $z/z_0 = 0\ldots 1.5$, the subsequent decay fits well to an exponential decay $\propto \exp(-z/z_{\rm{m}})$ with mean $z$-lifetime  $z_{\rm{m}}\approx 230~\mathrm{km}$ ($\approx 33 z_0$ with $z_0\approx 6.98~{\mathrm{km}}$), see Fig.~\ref{fig:S05a}(g).

\subsection{Multi-wavelet bound states}
\begin{figure}[t!]
\includegraphics[width=\linewidth]{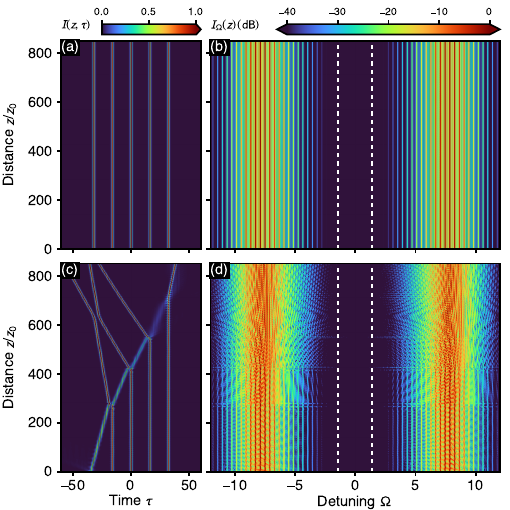}
\caption{Propagation dynamics of initial conditions consisting of sequences of 5 identical localized pulses.
(a) Initial condition Eq.~(\ref{eq:QC_ini}) governed by Eq.~(\ref{eq:HONSE}). All localized pulses have $\kappa\approx 18~\mathrm{km^{-1}}$, 
and the propagation distance is scaled by $z_0=2 \pi/\kappa$. Vertical dashed lines indicate zero-dispersion points.
(b) Same as (a) but with the leftmost localized pulse slightly offset towards its neighbor.
%
}
\label{fig:S06}
\end{figure}

The above binding strategy can be directly generalized to obtain multi-wavelet bound states, where, due to the effective exponential decrease of the wavelet intensity along its tails, only the nearest neighbor wavelet-wavelet interaction accounts for bound state formation \cite{Buryak:PRE:1995}.   
Figure~\ref{fig:S06}(a,b) demonstrates the propagation dynamics for a quasicrystal consisting of 5 wavelets, for which the initial condition is given by 
\begin{align}
A_0(\tau) = \sum_{n=-2}^{2} U(\tau + n\,\Delta\tau), \label{eq:QC_ini}
\end{align}
using the above wavelet $U$ as building block.
%
Following the discussion of the main document we demonstrate
that quasicrystals break up if neighboring wavelets are not properly interlocked at the local minima of $H_{\rm{min}}$: offsetting the leftmost wavelet by $\Delta \tau^\prime = -35.24~\mathrm{ps}$ instead of $-2 \Delta \tau = -32.04~\mathrm{ps}$ leads to the dynamics shown in Fig.~\ref{fig:S06}(c,d).
%

\section{Approximate Gaussian solutions}

As pointed out in the main document, as $\kappa \gg \max(D)$, the intensity of a wavelet-like solution is characterized by a prominent central peak. This suggests a further simplified description of solutions to Eq.~(\ref{eq:HONSE}) for $\beta_2>0$ and $\beta_4<0$, in terms of this peak alone. 
Following the local approach used to analyze pure quartic solitons in Ref.~\cite{Redondo:NC:2016}, we assume a Gaussian trial function and attempt to construct an approximate solution for Eq.~(\ref{eq:HONSE}) in terms of the Ansatz 
\begin{align}
A_{\rm{G}}(z,\tau)=A_0 \exp\{-(\tau/t_0)^2 + i\kappa z\}.\label{eq:G_ansatz}
\end{align} 
Based on the observation that solutions for large values of $\kappa$ appear to not strongly dependent on $\beta_3$, we below neglect the 3OD-term of Eq.~(\ref{eq:HONSE}).
Setting $q \equiv \tau/t_0$ then gives $\partial \tau = t_0^{-1} \partial_q$ and   
\begin{align}
i\partial_z A = c_1 \partial_q^2 A + c_2 \partial_q^4 A - \gamma |A|^2 A, \label{eq:HONSE_scaled} 
\end{align}
with $c_1\equiv |\beta_2|/[2\,t_0^2]$ and $c_2\equiv |\beta_4|/[24\,t_0^4]$. Evaluating the various partial derivatives for the Gaussian Ansatz~(\ref{eq:G_ansatz}) results in 
\begin{subequations}
\begin{align}
\partial_z\,A_{\rm{G}} &= i\kappa \,A_{\rm{G}}, \label{eq:G1}\\
\partial_q^2\,A_{\rm{G}} &= 2(2q^2-1)\,A_{\rm{G}} \equiv {\mathcal{H}}_2(q)\,A_{\rm{G}}, \label{eq:G2}\\
\partial_q^4\,A_{\rm{G}} &= (16 q^4 - 48q^2 + 12)\,A_{\rm{G}} \equiv {\mathcal{H}}_4(q)\,A_{\rm{G}},  \label{eq:G4}
\end{align}  
\end{subequations}
where ${\mathcal{H}}_n$ specifies the Hermite polynomial of order $n$. 
This allows to transform the partial differential equation Eq.~(\ref{eq:HONSE_scaled}) into the algebraic equation
\begin{align}
-\kappa \,A_{\rm{G}} = [c_1 {\mathcal{H}}_2(q) + c_2 {\mathcal{H}}_4(q)]\,A_{\mathrm{G}} - \gamma |A_{\mathrm{G}}|^2 A_{\mathrm{G}}. \label{eq:G5}
\end{align}

\begin{figure}[b!]
\includegraphics[width=\linewidth]{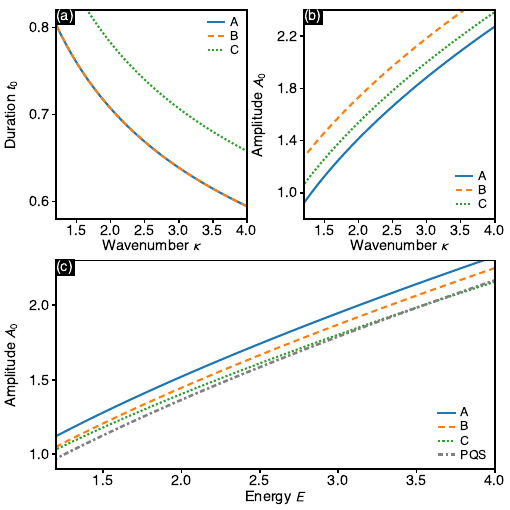}
\caption{Parameters of approximate Gaussian solutions for the three cases 
$(\beta_2,\beta_4)=(1~\mathrm{ps^2/km},-1~\mathrm{ps^4/km})$ (labeled A), 
$(\beta_2,\beta_4)=(0.5~\mathrm{ps^2/km},-1~\mathrm{ps^4/km})$ (labeled B), 
$(\beta_2,\beta_4)=(1~\mathrm{ps^2/km},-1.5~\mathrm{ps^4/km})$ (labeled C).
(a) Pulse duration, and,
(b) pulse amplitude as function of the wavenumber parameter $\kappa$.
(c) Dependence of the pulse amplitude on its energy. Dash dotted line indicates results for pure quartic solitons with $\beta_4=-1~\mathrm{ps^4/km}$.
}
\label{fig:S07}
\end{figure}

\noindent Restricted to small values of $q$, i.e.\ in the vicinity of the peak amplitude of Eq.~(\ref{eq:G_ansatz}), we expand $A_{\rm{G}}$ up to second order in $q$, giving
\begin{align}
A_{\mathrm{G}}&\approx A_0\,(1-q^2)\,e^{i\kappa z},\quad \text{and}, \label{eq:AG1}\\
|A_{\rm{G}}|^2A_{\rm{G}}&\approx A_0^3\,(1-3q^2)\,e^{i\kappa z}.\label{eq:AG2}
\end{align}
This allows to reduce the linear terms in Eq.~(\ref{eq:G5}) to
\begin{align}
c_1 {\mathcal{H}}_2(q) (1-q^2) &= 2 c_1 (-2q^4 + 3q^2 - 1),\quad\text{and}, \label{eq:G6}\\
c_2 {\mathcal{H}}_4(q) (1-q^2) &= 4 c_2 (-4q^6 + 16q^4 - 15q^2 + 3). \label{eq:G7}
\end{align}
Matching terms according to powers in $q$  yields for the two leading orders   
\begin{align}
q^0&:\quad \kappa = 2c_1 - 12 c_2 + \gamma A_0^2, \label{eq:cond_q0}\\
q^2&:\quad \kappa = 6c_1 - 60 c_2 + 3\gamma A_0^2. \label{eq:cond_q2}
\end{align}

\noindent Substituting Eq.~(\ref{eq:cond_q2}) into Eq.~(\ref{eq:cond_q0}) then produces an equation connecting the pulse duration $t_0$ and pulse amplitude $A_0$, given by
\begin{align}
t_0^2 = -\frac{|\beta_2|}{2 \gamma A_0^2} + \sqrt{ \frac{|\beta_2|^2}{4 \gamma^2 A_0^4} + \frac{|\beta_4|}{\gamma A_0^2}}. \label{eq:t0}
\end{align}
Finally, combining Eqs.~(\ref{eq:cond_q0}) and (\ref{eq:cond_q2}) to eliminate the $c_2$-term, provides an equation that relates the parameters of the pulse amplitude to the wavenumber in the form 
\begin{align}
\kappa = \frac{|\beta_2|}{2 t_0^2} + \frac{\gamma A_0^2}{2}. \label{eq:k}
\end{align}
When fixing a value of $\kappa$, Eqs.~(\ref{eq:t0}-\ref{eq:k}) can be cast into an implicit relation for the duration $t_0$, which can easily be solved by root-finding techniques.
%
The dependence of pulse duration and pulse amplitude on the wavenumber parameter, as well as the dependence of the energy $E=\int |A_{\rm{G}}(\tau)|^2~{\rm{d}}\tau=\sqrt{\pi/2}\,t_0\,A_0^2$ on the pulse amplitude is shown in Fig.~\ref{fig:S07}, comparing results for different choices of $\beta_2$ and $\beta_4$.
%
As can be expected from Eq.~(\ref{eq:t0}), changing the value of $\beta_2$ does not affect the resulting pulse duration by much, see the curves labeled A and B in Fig.~\ref{fig:S07}(a). 
Let us point out that, when letting $|\beta_2| \to 0$, the relations for the pulse duration and wavenumber match the expressions $t_0^2=\sqrt{|\beta_4|/(\gamma A_0^2)}$, and, $\kappa=\gamma A_0^2/2$ obtained earlier using a similar local approximation approach for pure quartic solitons~\cite{Redondo:NC:2016}.

\section{Impact of the Raman effect}

\begin{figure}[t!]
\includegraphics[width=\linewidth]{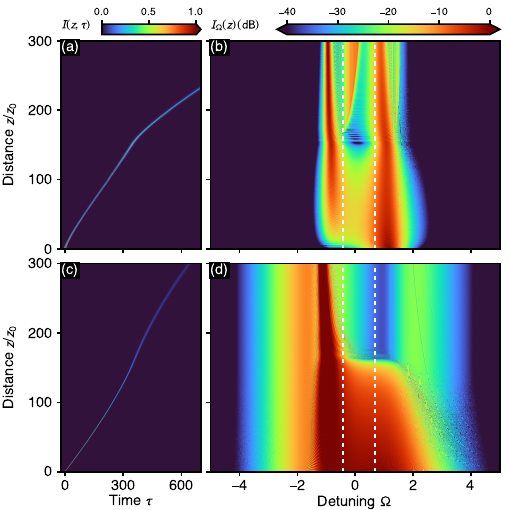}
\caption{Propagation dynamics of wavlets in presence of self-steepening and the Raman effect.
(a) Time-domain propagation dynamics, and, (b) corresponding spectrum for a wavelet with $\kappa=0.05~\mathrm{\mu m^{-1}}$.
(c,d) Same as (a,b) for a wavelet with $\kappa=0.30~\mathrm{\mu m^{-1}}$.
The propagation distance is scaled by $z_0=2 \pi/\kappa$. 
}
\label{fig:S08}
\end{figure}

As discussed in the main document, we performed pulse propagation simulations in terms of the generalized nonlinear Schrödinger equation (GNSE)
\begin{align}
i \partial_z A =&\frac{\beta_2}{2} \partial_\tau^2 A + i\frac{\beta_3}{6}\partial_\tau^3 A - \frac{\beta_4}{24} \partial_\tau^4 A - \gamma \left( 1+\frac{i\partial_\tau}{\omega_0}\right) \notag\\
&\times \left[\,A(z,\tau)\int R(\tau^\prime) |A(z,\tau-\tau^\prime)|^2~{\rm{d}}\tau^\prime\right],  \label{eq:GNSE}
\end{align}
with parameters $\gamma=10^{-10}~\mathrm{W^{-1}/\mu m}$, $\beta_2=0.1~\mathrm{fs^2/\mu m}$, $\beta_3=0.1~\mathrm{fs^3/\mu m}$, $\beta_4=-0.7~\mathrm{fs^4/\mu m}$, and reference frequency $\omega_0=2~\mathrm{(rad/fs)}$.
The Raman effect is included via $R(t)=(1-f_{\rm{R}})\,\delta(t) + f_{\rm{R}}\,h_{\rm{R}}(t)$, where the Raman response function
\begin{align}
h_{\rm{R}}(t) = \frac{\tau_1^2 + \tau_2^2}{\tau_1 \tau_2^2}\,e^{-t/\tau_2}\,\sin(t/\tau_1)\,\Theta(t), \label{eq:hR_t}
\end{align}
enters with fractional contribution $f_{\rm{R}}=0.2$, $\tau_1 = 12.2~\mathrm{fs}$, and $\tau_2=32~\mathrm{fs}$ \cite{Blow:JQE:1989}.

Figure~\ref{fig:S08} shows pulse propagation simulations of Eq.~(\ref{eq:GNSE}) in terms of the conservation quantity error method \cite{Heidt:JLT:2009,Melchert:CPC:2022}, using the conserved
quantity $C(z)=\sum |A_\Omega|^2/(\Omega+\omega_0)$ \cite{Blow:JQE:1989} to guide stepsize selection.
As initial conditions we use wavelets at selected wavenumbers $\kappa$, obtained for the HONSE~(\ref{eq:HONSE}).
For the small wavenumber $\kappa=0.05~\mathrm{\mu m^{-1}}$, the wavelet is localized at large frequencies [Fig.~\ref{fig:S08}(a,b)].
At the large wavenumber $\kappa=0.30~\mathrm{\mu m^{-1}}$, the wavelet stretches over the entire domain of normal dispersion. 
Both wavelets defy the usual SFS on short propagation distances, but are prone to spectral tunneling on long distances, with phase matching processes driving a transfer of energy from higher to lower frequencies across the domain of normal dispersion \cite{Serkin:EL:1993,Tsoy:PRA:2007}.


\bibliography{references}

\end{document}